# Phonon and Thermal Properties of Quasi-Two-Dimensional FePS₃ and MnPS₃ Antiferromagnetic Semiconductor Materials

**Fariborz Kargar[1,×,*], Ece Aytan[1,×], Subhajit Ghosh[1], Jonathan Lee[2], Michael Gomez[2], Yuhang Liu[3], Andres Sanchez Magana[1], Zahra Barani Beiranvand[1], Bishwajit Debnath[3], Richard Wilson[2,4], Roger K. Lake[3,4], and Alexander A. Balandin[1,4,*]**

[1]Phonon Optimized Engineered Materials (POEM) Center, Department of Electrical and Computer Engineering, University of California, Riverside, California 92521 USA

[2]Mechanical Engineering Department and Materials Science and Engineering Program, University of California, Riverside, California 92521 USA

[3]Laboratory for Terascale and Terahertz Electronics (LATTE), Department of Electrical and Computer Engineering, University of California, Riverside, California 92521 USA

[4]Spins and Heat in Nanoscale Electronic Systems (SHINES) Center, University of California, Riverside, CA 92521 USA

[×] Contributed equally to the work.
[*] Corresponding authors (A.A.B.): balandin@ece.ucr.edu and (F.K): fkargar@engr.ucr.edu;
web-site: http://balandingroup.ucr.edu/

## Abstract

We report results of investigation of the phonon and thermal properties of the exfoliated films of layered single crystals of antiferromagnetic $FePS_3$ and $MnPS_3$ semiconductors. The Raman spectroscopy was conducted using three different excitation lasers with the wavelengths of 325 nm (UV), 488 nm (blue), and 633 nm (red). The resonant UV-Raman spectroscopy reveals new spectral features, which are not detectable via visible Raman light scattering. The thermal conductivity of $FePS_3$ and $MnPS_3$ thin films was measured by two different techniques: the steady-state Raman optothermal and transient time-resolved magneto-optical Kerr effect. The Raman optothermal measurements provided the orientation-average thermal conductivity of $FePS_3$ to be $1.35 \pm 0.32 \, Wm^{-1}K^{-1}$ at room temperature. The transient measurements revealed that the through-plane and in-plane thermal conductivity of $FePS_3$ is $0.85 \pm 0.15 \, Wm^{-1}K^{-1}$ and $2.7 \pm 0.3 \, Wm^{-1}K^{-1}$, respectively. The films of $MnPS_3$ have higher thermal conductivity of $1.1 \pm 0.2 \, Wm^{-1}K^{-1}$ through-plane and $6.3 \pm 1.7 \, Wm^{-1}K^{-1}$ in-plane. The data obtained by both techniques reveal strong thermal anisotropy of the films and the dominant contribution of phonons to heat conduction. Our results are important for the proposed applications of the antiferromagnetic semiconductor thin films in spintronic devices.

# ■    INTRODUCTION

Transition-metal phospho-trichalcogenides, $MPX_3$, where M is a transition metal *e.g.* V, Mn, Fe, Co, Ni, or Zn and X is a chalcogenide as S, Se, Te, span a wide variety of layered compounds with different electronic, optical, and magnetic properties[1]. Bulk $MPX_3$ materials have been extensively studied during the last three decades mostly because of their potential application as cathodes in lithium batteries[2,3]. With the beginning of the era of two−dimensional (2D) materials after exfoliation of stable atomically thin single layer graphene[4] and discovery of its unique electronic[5–7] and thermal properties[8,9], the attention turned to quasi-2D films of transition-metal dichalcogenides ($MX_2$) and $MPX_3$. It has been demonstrated that some $MX_2$ exhibit ferromagnetic (FM) properties at monolayer thicknesses even at room temperature (RT)[10,11]. Motivated by the desire to find intrinsic antiferromagnetic (AFM) ordering in the 2D limit, it was discovered that $MPX_3$ structures are one of those rare few-layer van der Waals (vdW) materials, which can have stable intrinsic antiferomagnetism even at mono- and few layer thicknesses[12].

The existence of weak vdW bonds between the $MPX_3$ layers makes them a potential candidate for the 2D spintronic devices. The "cleavage energy" of these materials − the energy required to separate a crystal into two parts along a basal plane − is close to that of graphite[13]. More specifically, the "cleavage energy" of $FePSe_3$ is slightly higher than that of the graphite while that for all other combinations of the M and X elements is lower than that of graphite. The Néel temperature, $T_N$, for $FePS_3$, $MnPS_3$, and $NiPS_3$ is reported to be around 118 K, 78 K, and 155 K, respectively[14]. The M element determines the type of the phase transition from AFM to paramagnetic (PM) ordering. While $FePS_3$ shows Ising-type phase transition at $T_N$, $MnPS_3$ and $NiPS_3$ follow Heisenberg- and XY-phase transitions, respectively[12]. The metal element of the $MPX_3$ materials modifies the band gap from a medium band gap of ~1.3 eV to a wide band gap of ~3.5 eV suitable for optoelectronic applications [13,14]. While $FePS_3$ has an indirect bandgap of 1.5 eV, $MnPS_3$ exhibits a direct bandgap of 3.0 eV, respectively[2]. The diverse properties of these materials tunable by proper selection and combination of the M and X elements make the $MPX_3$ materials an interesting platform for fundamental science and practical applications in spintronic devices[15,16], lithium batteries[17], field-effect transistors[18], UV light detectors[18], thermoelectrics[19], and photocatalytic systems[20]. The semiconductor nature of $FePS_3$ and $MnPS_3$ and the possibility of electron and phonon band-structure engineering with strain[14] make these materials particularly interesting from the fundamental and practical applications points of view.

The optical phonon properties of $FePS_3$ and $MnPS_3$ bulk crystals have been studied extensively using Raman spectroscopy and infra-red (IR) absorption techniques[12–14,21–31]. The early Raman studies of bulk crystals suggested that the vibrational dynamics of both crystals can be pictured within the framework of a molecular approach, dividing the Raman spectrum into the high-frequency internal modes of $P_2S_6$, and the low-frequency modes where the interaction of the M element with phosphorous (P) and chalcogenides (X) becomes significant. A recent temperature-dependent polarized Raman study of bulk, few- and monolayer $FePS_3$ revealed new features in the spectra, emerging at the transition temperature, and suggested that a monolayer $FePS_3$ possesses an Ising-type AFM ordering[12]. Another Raman study of $MnPS_3$ indicated that $MnPS_3$ exhibits three phase transitions at 55 K, 80 K, and 120 K, which were attributed to unbinding of spin vortices, transition from an AFM to a PM state, and two-dimensional spin critical fluctuations, respectively[23]. Although, available Raman and IR-absorption studies are in general agreement, there exist ambiguities with peak assignments in the frequency range from 400 cm$^{-1}$ to 500 cm$^{-1}$.

Some IR[29] and Raman[26] investigations reported an anomalous peak at ~480 cm$^{-1}$. While this peak was assigned to the normal vibration of the P–P bond the available theoretical reports[32] do not show any optical phonon modes at this frequency. Other experimental studies found the P–P mode peak at ~430 cm$^{-1}$ as a weak feature in Raman and strong peak in IR-absorption spectroscopy. Since the peak at 480 cm$^{-1}$ exhibits low intensity, and it emerges as a broad shoulder in the visible laser Raman and IR spectra, the nature of these spectral feature has been questioned. In the present study, we report the vibrational dynamics of crystalline thin films of FePS$_3$ and MnPS$_3$ using three different lasers, with the excitation wavelengths of 325 nm (UV), 488 nm (blue), and 633 nm (red). The Raman spectra obtained by blue and red lasers are in excellent agreement with the available literature. The data acquired by UV-Raman spectroscopy are similar for both crystals, revealing an intense peak at ~467 cm$^{-1}$. We also report the temperature dependent and excitation power dependent coefficients for the phonon modes in both material systems. The coefficients are required for understanding the anharmonicity of the crystal lattice and for determining the average thermal conductivity using the optothermal Raman technique.

Although magnetic and electronic properties of MPX$_3$ family compounds have been investigated intensively, there have been no experimental reports on the thermal properties of these materials. Like other layered crystalline materials, the members of the MPX$_3$ family are expected to have strongly anisotropic thermal properties owing to their weak vdW interlayer bonds and strong in-plane covalent bonds. A recent theoretical study reported an exceptionally low thermal conductivity of mono-layer ZnPSe$_3$, which makes it a potential candidate for thermoelectric applications[19]. The knowledge of thermal transport properties of quasi-2D MPX$_3$ materials is important for spintronic and thermoelectric applications[16]. It has been suggested that AFM materials have advantages over FM materials in spintronic applications, *e.g.* in spin-Seebeck effect (SSE) devices, since they are less susceptible to external magnetic fields, and possess a linear magnon dispersion in the vicinity of Brillouin zone (BZ) center with a high group velocity[16]. In the AFM based SSE devices[15], the AFM material can be utilized, instead of FM material, as the spin generation layer. The generated voltage, $V$ by the SSE device, as a result of the induced temperature gradient, $\Delta T$, by a power source, $P$, across the AFM layer is inversely proportional to its thermal conductivity, $k$, *i.e.* $V \propto \Delta T \propto P/k$. It has been theoretically suggested that polar AFM materials like $MnPS_3$ with honeycomb structure and broken inversion symmetry, can be used in SSE devices to act as a nonreciprocal spin transport medium[33]. For these reasons, the knowledge of the thermal properties of $MPX_3$ materials becomes essential for designing the spintronic devices. In this study, we use the steady-state optothermal Raman and transient time-resolved magneto-optical Kerr effect (TR-MOKE) techniques in order to determine the direction-average, in-plane, and through-plane thermal conductivity values of the exfoliated thin films of the high-quality $FePS_3$ and $MnPS_3$ single crystals.

■    **EXPERIMENTAL SECTION**

**Materials and characterization:** Commercially available high-quality single crystals of $FePS_3$ and $MnPS_3$ (HQ Graphene) were used for Raman spectroscopy and TR-MOKE experiments. The materials were synthesized using the chemical vapor transport (CVT) method. Figure 1 (a-b) shows the results of X-ray diffraction spectroscopy (XRD) of $FePS_3$ and $MnPS_3$ crystals, respectively. The XRD data reveals that both materials possess a pure single crystal phase with major (001) and (002) crystallographic planes[21]. The insets in Figure 1 (a-b) illustrate the crystal structure of both material systems where the grey, yellow, gold, and violate spheres represent phosphorous (P), sulfur (S), iron (Fe), and manganese (Mn) atoms, respectively. The red arrows show the direction of the spin ordering in metallic atoms in each layer. The energy-dispersive X-ray spectroscopy (EDX) data confirming the purity of chemical structures of the crystals is presented in Supplementary Figure S1.

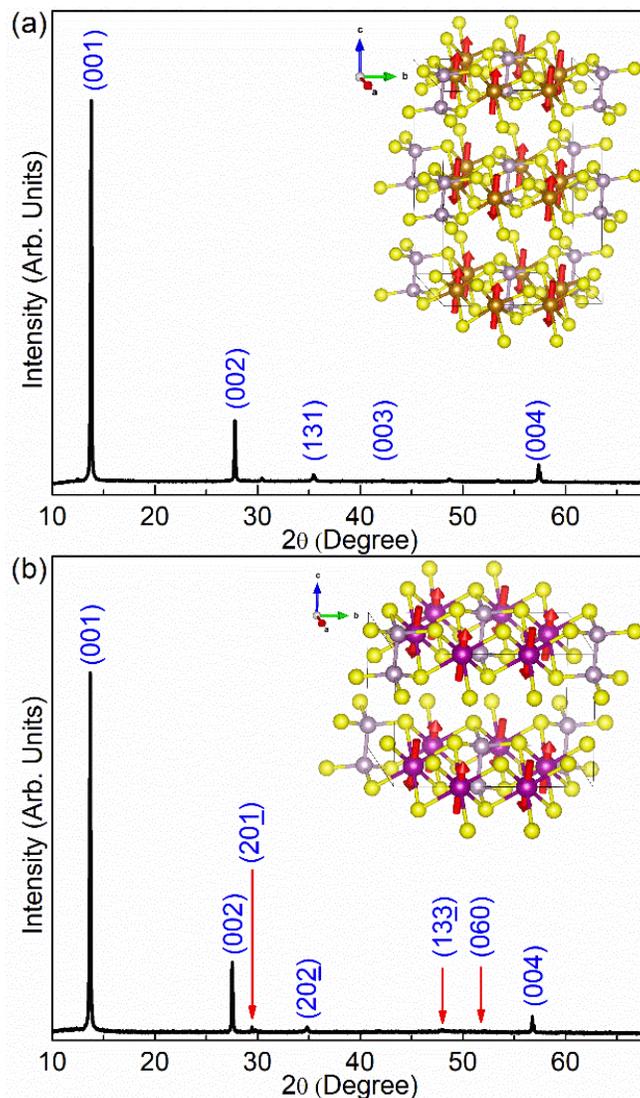

**Figure 1:** X-ray diffraction (XRD) patterns of high-quality single crystals of (a) $FePS_3$ and, (b) $MnPS_3$. Both spectrums exhibit intense peaks corresponding to the crystallographic planes of (001) and (002) confirming the high-quality single crystal structure of the material systems. The insets in (a) and (b) demonstrate the atomic structure of the $FePS_3$ and $MnPS_3$ where the grey, yellow, gold, and violate spheres represent phosphorous (P), sulfur (S), iron (Fe), and manganese (Mn) atoms, respectively. The arrows show the spin ordering of both crystals in antiferromagnetic state.

The crystals were mechanically exfoliated onto $Si/SiO_2$ and diamond substrates. The resulting multilayer flakes had lateral dimensions of 5 μm to 25 μm. The optical images of the representative flakes are shown in Supplementary Figure S2. The thicknesses of the samples were verified with the atomic force microscopy (AFM) (Supplementary Figure S3). Raman spectroscopy (Renishaw inVia) measurements were performed in the backscattering configuration using three lasers with different excitation wavelengths of 325 nm (UV), 488 nm (blue), and 633 nm (red). The detector cut-off frequencies for these wavelength were 300 cm$^{-1}$, 10 cm$^{-1}$, and 100 cm$^{-1}$, respectively. In order to avoid self-heating effects, we used law laser excitation power in order to avoid self-heating effects.

■    **RESULTS AND DISCUSSION**

The room temperature (RT) Raman spectra of FePS$_3$ and MnPS$_3$ on silicon (Si) substrate at three different laser excitations are presented in Figure 2 (a-b). The corresponding wavenumbers of the observed peaks are listed in Table 1. In all spectra, the sharp peak at 520 cm$^{-1}$ is originated from the silicon substrate. One can notice similarities in the Raman normal modes for both materials, at each laser excitation, except for the lowest normal modes of spectra excited by the blue laser. Although the MPX$_3$ structures have a complex atomic configuration, their Raman signatures can be divided into the internal vibrational modes of the ethane-type $P_2S_6^{4-}$ and external vibrational modes of the $M^{2+}$ and $P_2S_6^{4-}$ [Ref. [27]]. The higher frequency modes in the spectra of both materials originate from the internal modes of $P_2S_6^{4-}$. Since this is common to both crystals, the associated Raman modes are nearly identical. The bulk MPX$_3$ crystals belong to the $C_{2/m}$ space symmetry group with the irreducible representation of $\Gamma = 8A_g + 6A_u + 7B_g + 9B_u$ zone-center modes, in which only the $A_g$ and $B_g$ modes are active in the regular Raman spectroscopy with the blue and red laser excitation[27,31]. All Raman peaks obtained under red and blue laser excitations are in excellent agreement with available literature[21–31]. The Raman spectra obtained under UV laser excitation shows two new peaks for both crystals, identified by a red star in Figure 2 (a-b), which are not detectable by visible light Raman spectroscopy.

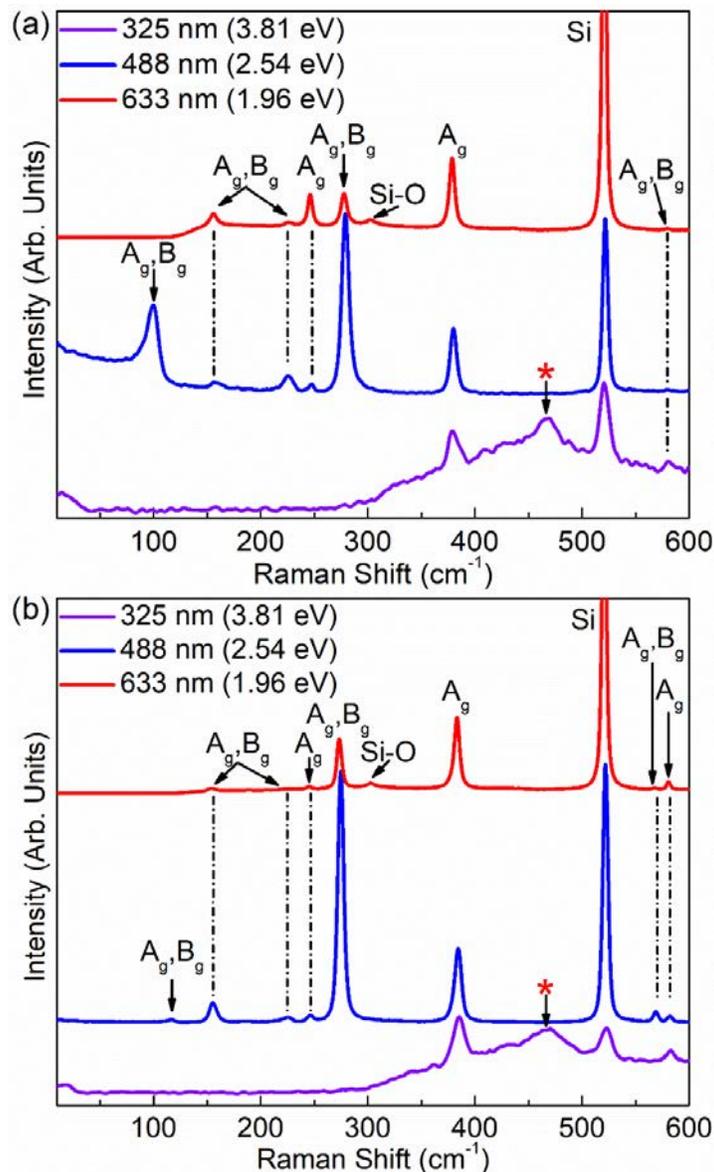

**Figure 2:** Raman spectroscopy of (a) FePS$_3$ and (b) MnPS$_3$ single crystals at room temperature using three different laser excitations of 325 nm (UV), 488 nm (blue), and 633 nm (red). The results of UV laser excitations exhibit a unique combined peak at ~465 cm$^{-1}$ for both crystals which have not been reported previously.

**Table 1:** Raman normal modes of $FePS_3$ and $MnPS_3$ at different laser excitation wavelengths

| | Excitation Laser | Raman Normal Modes ($cm^{-1}$) | | | | | | | | | |
|---|---|---|---|---|---|---|---|---|---|---|---|
| | | $A_g, B_g$ | $A_g, B_g$ | $A_g, B_g$ | $A_g$ | $A_g, B_g$ | $A_g$ | $B_u$ | $B_g$ | $A_g, B_g$ | $A_g$ |
| $FePS_3$ | UV | – | – | – | – | – | 379 | 431 | 467 | – | 580 |
| | Blue | 98 | 158 | 226 | 247 | 279 | 380 | – | – | – | 579 |
| | Red | – | 156 | 227 | 246 | 278 | 378 | – | – | – | 579 |
| $MnPS_3$ | UV | – | – | – | – | – | 385 | 428 | 467 | – | 582 |
| | Blue | 117 | 156 | 225 | 246 | 274 | 384 | – | – | 568 | 581 |
| | Red | – | 156 | 226 | 245 | 273 | 383 | – | – | 567 | 581 |

The UV-Raman spectra of $FePS_3$ and $MnPS_3$ on Si are shown separately in Figure 3 (a-b), respectively. There are four peaks identified as $P_i, i = 1 \ldots 4$ in this Figure. The peaks marked as $P_1$ and $P_4$ are present in the Raman spectra excited by red and blue lasers (Figure 2). The peak $P_2$ at ~430 $cm^{-1}$ is not detectable by visible Raman spectroscopy. Under UV laser excitation, this spectral feature appears as a low-intensity shoulder. A peak at the same frequency has been reported previously using IR spectroscopy[29]. It was attributed to the zone-center phonon mode with $B_u$ symmetry which is not a regular Raman active mode. There is also a distinguishable broad peak, labeled as $P_3$, which is a unique feature detectable only with the UV Raman spectroscopy. As one can see in Figure 3, the $P_2$ and $P_3$ features can be fitted accurately using two separate Gaussian peaks (green curves). The spectral position of each individual constituent of the $P_3$ feature is almost identical for both materials (Table 1). Comparing the experimental UV-Raman results with the available theoretical phonon dispersion data[32], one notices that there is no phonon band, either at the BZ center or at the high-symmetry points of the BZ boundaries, corresponding to the frequencies associated with $P_3$. We note that somewhat similar spectral features have been reported using IR spectroscopy[29] as low intensity peaks defined as shoulders. Since the experiments were conducted at RT, which is significantly higher than the Néel temperature for both materials, it is unlikely that this spectral feature is associated with the light scattering by one- or two-magnons. The UV-Raman spectrum of a bare Si substrate, and on different spots of various exfoliated flakes are presented in Supplementary Figure S4, Figure S5, and Figure S6 confirming that the observed $P_3$ peak originates form the MPX$_3$ crystals themselves. Prior theoretical studies[32] report that there should be a Raman active normal mode for both crystals at 520 cm$^{-1}$. However, as shown in Figures 2 (a-b), the silicon substrate also shows an intense peak at 520 cm$^{-1}$ which masks Raman features originating from the MPX$_3$ crystals. To address this issue, UV-Raman measurements were conducted on FePS$_3$ flakes exfoliated on a diamond substrate, and the results are shown in Figure 3 (c). Comparing the Raman intensity from the bare diamond substrate (orange curve) and FePS$_3$ on diamond (violet curve) confirms that there are no vibrational modes at 520 cm$^{-1}$ associated with FePS$_3$. The UV-Raman spectroscopy of bare diamond substrate is shown in Supplementary Figure S7.

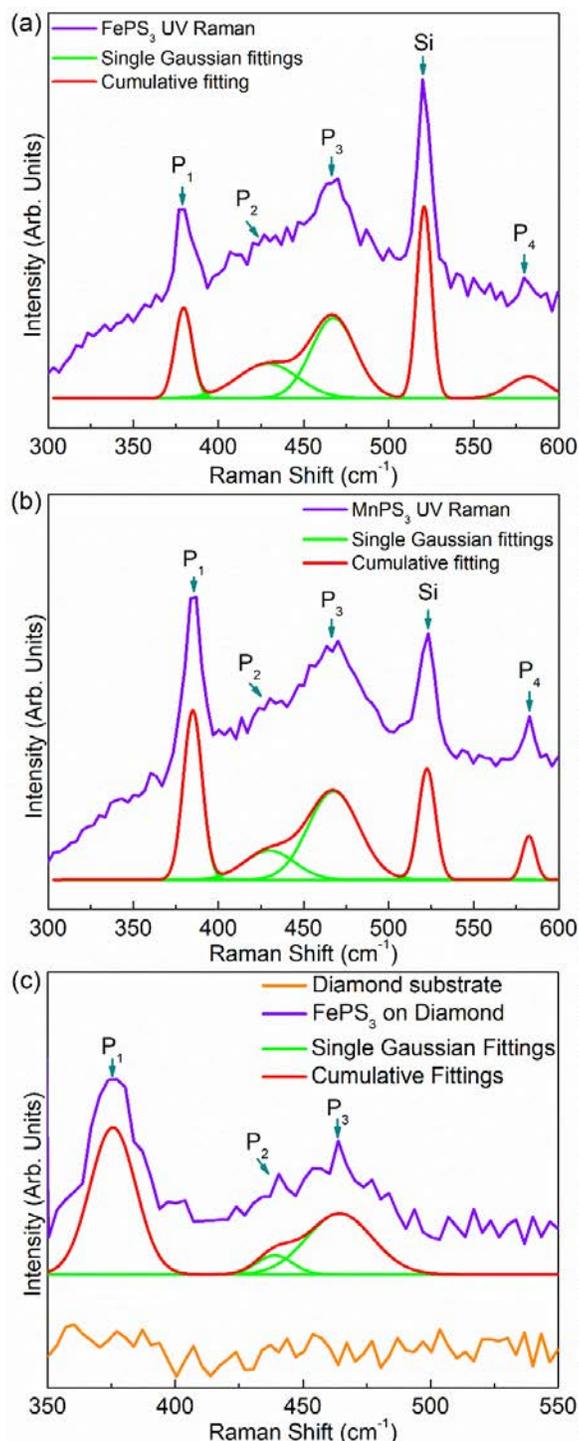

**Figure 3:** Raman spectroscopy of (a) FePS$_3$, (b) MnPS$_3$ on silicon substrate, and (c) FePS$_3$ on diamond substrate at room temperature using UV laser excitation wavelength. The violate curves are the experimental data. The green and red curves represent individual Gaussian and cumulative fittings to experimental data, respectively. The shoulder and peak identified as P$_2$ at ~430 cm$^{-1}$ and P$_3$ at ~465 cm$^{-1}$at are unique features which are only observable with UV laser excitation. The rest of the modes are detectable by red and blue laser excitation Raman spectroscopy as well. The peak marked as Si is associated with the silicon substrate.

We performed density functional theory calculations of bulk $FePS_3$ as implemented in VASP[34,35] to understand the nature of the two experimentally observed modes between 400 and 500 cm[-1]. Exchange-correlation is included with the Perdew-Burke-Ernzerhof (PBE) functional[36]. A Hubbard-U type correction is applied to the d-orbitals of the Fe atoms using the VASP default model that depends only on the value of *U-J*. Van der Waals forces are included with the DFT-D2 method of Grimme[37]. Atomic positions are relaxed until the Hellman-Feynman forces are less than $10^{-4}$ eV/Å. Total energies are converged to within $10^{-7}$ eV in both the geometry relaxation and the phonon calculations. Full details of the calculations are provided in the Supplementary Materials. We find that the calculated frequencies depend on the spin texture, the supercell, and the particular parameters used. The vibrational modes are calculated for a structure in the zigzag AFM (z-AFM) phase and for a structure in the non-magnetic phase. The z-AFM phase is the low-temperature ground state. Two different calculations are performed for the z-AFM phase. The first calculation uses a 2×2×2 supercell and *U-J* = 5.91 eV. The second calculation uses a 4×4×1 supercell and *U-J* = 3.5 eV. The calculation of the non-magnetic phase uses a 4×4×1 supercell and *U-J* = 3.5 eV. The resulting frequencies are given in Table 2. Both the non-magnetic and the z-AFM 4×4×1 supercells show a Raman inactive mode at 417 cm[-1], and the z-AFM 2×2×2 supercell shows a Raman inactive mode at 431 cm[-1]. The higher frequency Raman active mode shows more variation. In the 2×2×2 z-AFM and the 4×4×1 non-magnetic supercells, the frequencies are 521 and 526 cm[-1], respectively, which are similar to frequencies previously calculated by others. The 4×4×1 z-AFM supercell gives a frequency of 483 cm[-1]. The atomic displacements of the Raman active mode are the same for all 3 simulations. Only the frequency changes. The displacements for the $B_u$ and $B_g$ modes are shown in Figure 4. Thus, the frequencies of these two modes are sensitive to the spin texture of the Fe atoms and the model parameters, and they appear to be the possible candidates for the origin of the experimentally observed modes at 431 and 467 cm$^{-1}$.

**Table 2.** Calculated frequencies and symmetries of the two modes closest to the experimentally observed modes at 431 and 467 cm$^{-1}$ from 3 different calculations

| Wavenumber (cm$^{-1}$) | Frequency (THz) | Irreducible representation | Raman |
|---|---|---|---|
| 4×4×1 supercell z-AFM FePS$_3$ | | | |
| 417 | 12.5 | $B_u$ | Inactive |
| 483 | 14.4 | $B_g$ | Active |
| 2×2×2 supercell z-AFM FePS$_3$ | | | |
| 431 | 12.9 | $B_u$ | Inactive |
| 521 | 15.6 | $B_g$ | Active |
| 4×4×1 supercell non-magnetic FePS$_3$ | | | |
| 417 | 12.5 | $B_u$ | Inactive |
| 526 | 15.8 | $B_g$ | Active |

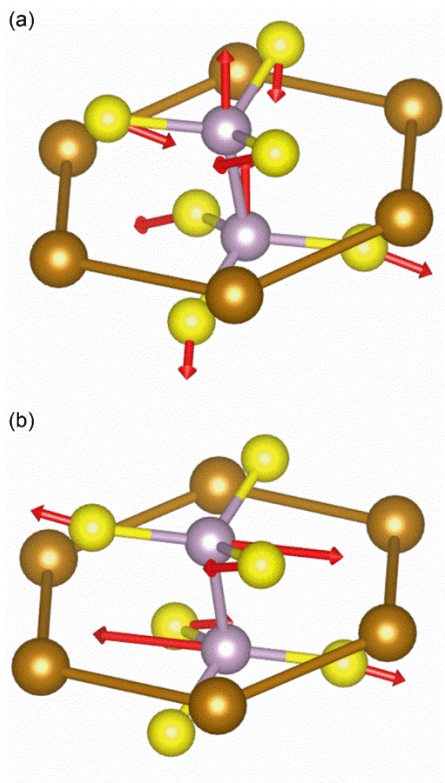

**Figure 4:** Calculated atomic displacements. (a) Lower frequency Raman inactive mode in frequency range 416-431 cm$^{-1}$. (b) Higher frequency Raman active mode in frequency range 483-526 cm$^{-1}$.

We investigated the evolution of Raman spectra of FePS$_3$ and MnPS$_3$ with temperature. The information on temperature coefficients of Raman modes can be used in the analysis of the crystal lattice inharmonicity and for extraction of the thermal data in the optothermal Raman method. This method has been widely used for measurements of the thermal conductivity of different free-standing and supported 2D materials[8,38–41]. It uses the Raman spectrometer both as a local heating element and as a thermometer. The measurement procedure involves three steps: temperature dependent Raman measurements, excitation power-dependent measurements, and data extraction via iterative solution of the inverse heat diffusion problem. For the temperature dependent Raman measurements, the sample is mounted in a hot-cold stage (Linkam Scientific, THMS-600), where its temperature can be controlled externally. The Raman measurements are conducted with the lowest possible laser excitation power at each individual sample temperature. The low laser power is essential at this step to avoid self-heating effects. However, it should be sufficient to achieve a reasonable signal-to-noise (S/N) ratio in the accumulated spectra.

The spectral positions of $A_g$ Raman peaks of FePS$_3$ and MnPS$_3$ as the function of temperature, from RT to 350 K, are plotted in Figure 5 (a-b), respectively. The Raman spectra used to plot the dependence in Figure 5 (a-b) are provided in Supplementary Figure S8 (a-b). For both crystals, the increase in temperature results in shifting of the $A_g$ phonon peaks to lower frequencies. We used a linear fit to determine the Raman temperature coefficient, $\chi = d\omega/dT$. Here, $\omega$ is the frequency of the Raman $A_g$ mode and $T$ is the sample's temperature, respectively. One can see an excellent linear fit of the $A_g$ Raman mode over the examined temperature range. The extracted Raman temperature coefficients are $\chi_{\text{Fe}} = -0.0154 \text{ cm}^{-1}\text{K}^{-1}$ and $\chi_{\text{Mn}} = -0.0108 \text{ cm}^{-1}\text{K}^{-1}$ for FePS$_3$ and MnPS$_3$, respectively. One should note that the Raman temperature coefficients are different for each vibrational mode. Their value depends on the intrinsic material properties and the examined temperature range. In the optothermal Raman technique, the extracted thermal conductivity is independent of the specific peak chosen for the optothermal Raman analysis. In general, an intense narrow phonon peak with pronounced temperature dependence is preferred. In our case, the $A_g$ Raman peaks of FePS$_3$ and MnPS$_3$ with Raman spectral positions of ~378 cm$^{-1}$ and ~383 cm$^{-1}$ at RT satisfy these requirements.

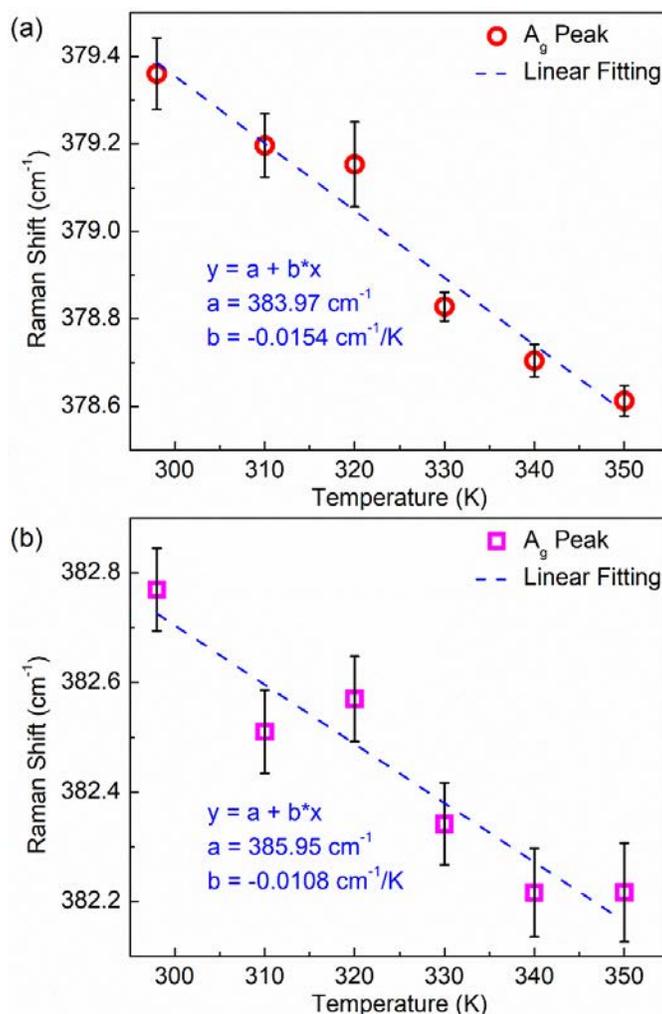

**Figure 5:** Raman A$_{1g}$ peak of (a) FePS$_3$ and (b) MnPS$_3$ as a function of the sample temperature. The measurements were conducted at very low laser excitation power to avoid local heating. The peaks for both crystals exhibit a linear decrease in Raman shift with increasing the sample temperature. The obtained dependence is used as a calibration curve for the thermal conductivity measurements via optothermal Raman method.

The next step of the optothermal Raman measurements is to determine the spectral position of the $A_g$ Raman mode as a function of increasing excitation laser power. The measurements were conducted at 633 nm wavelength (red laser). The laser power on the sample surface at each step was measured with a power meter (Newport 843-R). Increasing the excitation laser power results in the local laser-induced heating of the sample and the corresponding shift of the frequency of the Raman modes due to the local temperature rise. Figure 6 (a-b) shows the measured $A_g$ Raman shifts of FePS$_3$ and MnPS$_3$ as a function of the laser incident power, respectively. The $A_g$ Raman mode of FePS$_3$ reveals a clear linear decrease in frequency with increasing the laser power, while for MnPS$_3$, the data are scattered. The power dependent Raman results depend on the optical properties of materials, *e.g.* reflection and absorption coefficients of the material at the given laser wavelength. The absorption coefficient of MnPS$_3$ at the excitation laser wavelengths of 488 nm and 633 nm is less than 1200 cm$^{-1}$ [Ref.[2]], which indicates that most of the light is not absorbed by the material, and thus, the sample experiences minor local heating by the incident laser light. For FePS$_3$, however, the absorption coefficient is ~12000 cm$^{-1}$ [Ref.[2]], almost ten times of that of MnPS$_3$, and therefore, the sample is heated strongly, exhibiting a clear Raman red-shift of the $A_g$ phonon mode. The change in the frequency of the FePS$_3$ Raman $A_g$ mode as a function of the incident laser power on the sample's surface ($P_D$) is defined as $\theta = d\omega / dP_D$ and found to be $\theta_{Fe} = -0.6723 \text{ cm}^{-1}\text{mW}^{-1}$. The weak dependence of the MnPS$_3$ $A_g$ Raman mode on laser power prevents its use for extracting thermal conductivity in the optothermal Raman method.

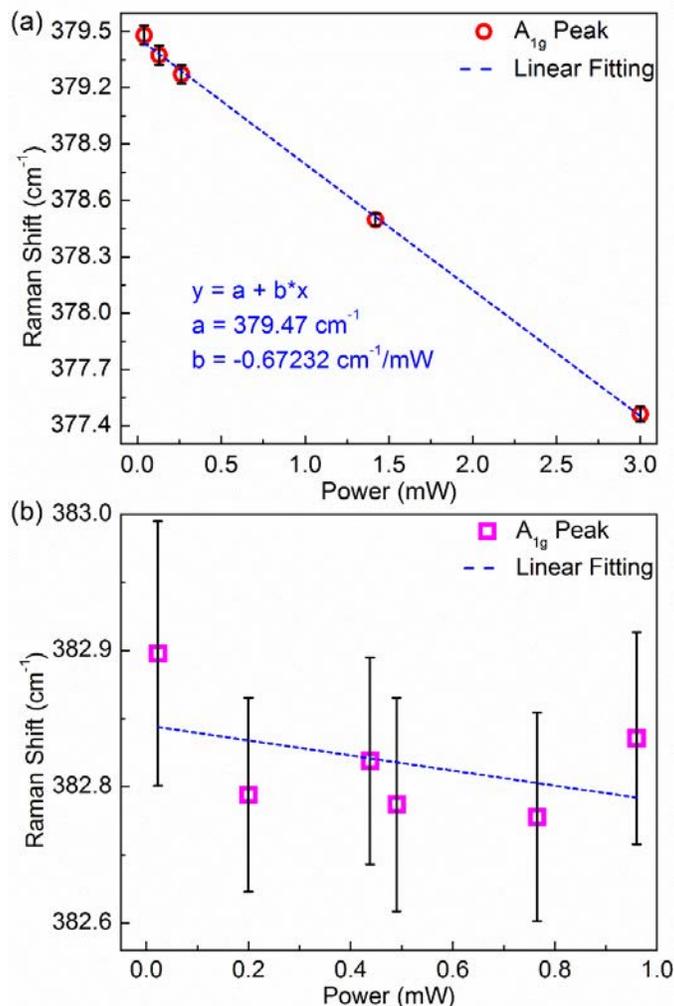

**Figure 6:** Raman $A_{1g}$ peak of (a) $FePS_3$ and (b) $MnPS_3$ as a function of the laser excitation power. The samples were initially at room temperature. For $FePS_3$, the $A_{1g}$ peak decrease linearly with increasing the laser power as a result of local heating. In $MnPS_3$, however, the change in Raman shift is within the errors of measurement due to the low absorption coefficient of MnPS3 at laser excitation wavelength which prevents local heating of the sample.

The final step of the Raman optothermal method is determining the thermal conductivity by solving the inverse heat-diffusion problem for the known geometry of the sample. The lateral dimensions and thickness of $FePS_3$ are determined from optical and AFM inspection, respectively (Supplementary Figure S2 and Figure S3). The details of the thermal conductivity calculations are given in the Methods part. Layered materials, such as $FePS_3$ and $MnPS_3$, have weak van der Waals forces between the layers along the "c" direction and strong covalent bonds among the in-plane

"a" and "b" directions (the inset in Figure 1 (a)). They are expected to have larger in-plane thermal conductivity ($k_\parallel$) and lower through-plane thermal conductivity ($k_\perp$). The Raman optothermal method used with the relatively thick or bulk samples provides the directional average thermal conductivity, which has both in-plane and through-plane components. The thermal conductivity extraction procedure for $FePS_3$ is presented in Figure 7. A thermal conductivity of $1.35 \pm 0.32$ $Wm^{-1}K^{-1}$ was extracted for $FePS_3$ thin film with the thickness of 400 nm. Given the thickness range and smooth interface of the crystal one can assume that the phonon – boundary scattering does not play a significant role, and the extracted value can be considered as the average "bulk" thermal conductivity of $FePS_3$. The experimental uncertainty in the measured value is mostly associated with the uncertainty in determining the excitation laser spot size. It should be emphasized that the extracted thermal conductivity value is a weighted arithmetic mean of the through-plane ($k_\perp$) and in-plane thermal ($k_\parallel$) conductivities of $FePS_3$ as $k = \alpha k_\perp + \beta k_\parallel$ where $\alpha$ and $\beta$ are the weight coefficients and thus $\alpha + \beta = 1$. Based on the lateral dimensions of the flake size (>10 µm) and the laser spot size (~1.5 µm), the extracted value by the optothermal Raman technique reflects more of the through-plane thermal conductivity rather than the in-plane thermal conductivity, *i.e.* $\alpha > \beta$. This is confirmed by measuring the in-plane and through-plane thermal conductivities of both $FePS_3$ and $MnPS_3$ using TR-MOKE method as described below.

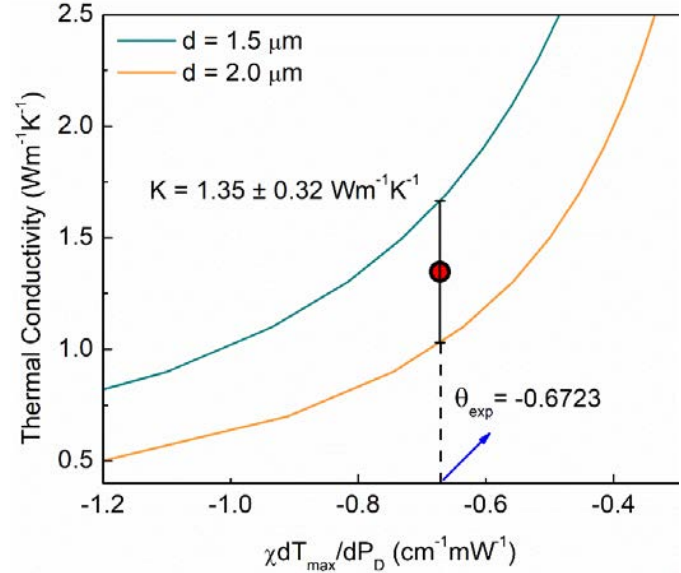

**Figure 7:** Numerical simulation of thermal conductivity of FePS$_3$ as a function of $\theta = \chi dT_{max}/dP_D$ using two different laser spot sizes. The dashed line presents $\theta$ as measured for FePS$_3$. The error bars exhibit the uncertainty in thermal conductivity calculations due to the uncertainty in laser spot size measurements. The extracted thermal conductivity is a weighted average of in-plane and cross-plane thermal conductivities of the crystal.

We determined separately the through-plane and in-plane thermal conductivity of the MnPS$_3$ and FePS$_3$ crystals with TR-MOKE measurements[42–44]. To facilitate the TR-MOKE measurements, we sputter coated the crystals with a ~15 nm thick Pt and Co multilayer thin-film. The geometry of the multilayer was Ta (3 nm) / Pt (3 nm) / [Co (0.8 nm)/Pt (1 nm)]x4/Pt (2 nm). We optically heated the metal film with a train of laser pulses that are modulated with a 50% duty cycle at frequency $f_{mod}$. The resulting temperature evolution of the sample surface was determined from the temperature-induced changes in the polar Kerr rotation of a reflected probe beam[45,46]. The rate of thermal transport out of the Pt-Co thin-film depends strongly on the thermal interface conductance between the crystal and the metal, as well as the thermal conductivity tensor of the MnPS$_3$ or FePS$_3$ crystals. The TR-MOKE method is analogous to the more established technique known as the time-domain thermoreflectance (TDTR)[42–46]. We chose to use TR-MOKE instead of TDTR to avoid coating the MnPS$_3$ and FePS$_3$ crystals with thick metal films. TDTR measurements require the sample of interest to be coated with an optically thick film, *e.g.* 80 nm of Al. TR-MOKE experiments do not require an optically opaque film[45,46]. Using a thick metal film as a transducer in the current experiments would shunt the in-plane heat current and reduce our measurement's sensitivity to the in-plane thermal conductivity of $FePS_3$ and $MnPS_3$. We use an analytical solution to the heat diffusion equation in cylindrical coordinates[42] to extract the thermal conductivity of the $MnPS_3$ and $FePS_3$ crystals, as well as the interface conductance ($G$) between the bottom Ta layer of the metal film and the $MnPS_3$ and $FePS_3$ crystals. Further details of our experimental method and setup are described elsewhere[42–44].

By performing measurements at multiple pump modulation frequencies, we can independently determine all three unknown thermal properties of the sample stack: $G$, $k_\perp$, and $k_\parallel$. Here, $k_\perp$ and $k_\parallel$ represent the through-plane and average in-plane thermal conductivity of the crystal in axial and radial directions in cylindrical coordinates, respectively. We assume the in-plane thermal conductivity is the same in all radial directions. For each crystal, measurements were performed with $f_{mod} = 1\ MHz$ and $10\ MHz$. In Figure 8, we plot our experimental data and the predictions of our thermal model. We find that the through-plane thermal conductivity of $FePS_3$ and $MnPS_3$ is $0.85 \pm 0.15\ \mathrm{Wm^{-1}K^{-1}}$ and $1.1 \pm 0.2\ \mathrm{Wm^{-1}K^{-1}}$, respectively. The in- plane conductivity of $FePS_3$ and $MnPS_3$ is $2.7 \pm 0.3\ \mathrm{Wm^{-1}K^{-1}}$ and $6.3 \pm 1.7\ \mathrm{Wm^{-1}K^{-1}}$, respectively. The interface conductance between the Ta seed-layer at the bottom of the metal multilayer and the $FePS_3$ and $MnPS_3$ crystals is $24 \pm 4\ \mathrm{MW\,m^{-2}K^{-1}}$ and $23 \pm 4\ \mathrm{MW\,m^{-2}K^{-1}}$, respectively. The interface provides an equivalent thermal resistance as ~60 nm of amorphous glass. Thermally resistive interfaces between metals and 2D materials are common due to phonon focusing in the in-plane direction of the 2D material[44]. The obtained thermal conductivity values for FePS₃ and MnPS₃ and layered materials of MX₃ and MX₂ families are summarized in Table 3.

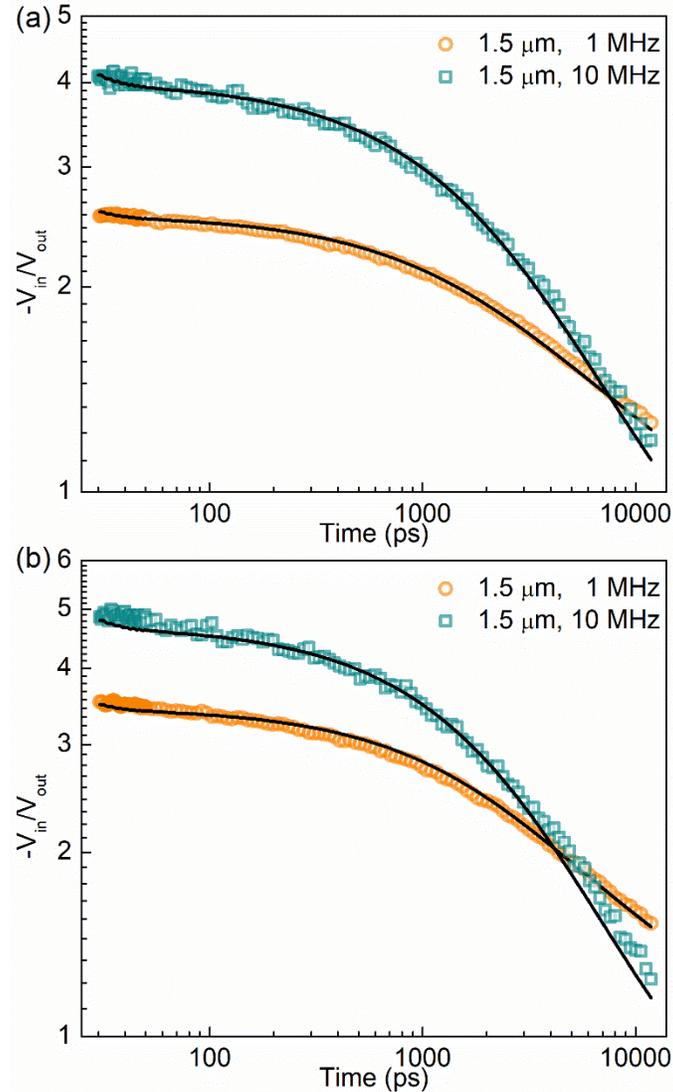

**Figure 8:** Time-resolved magneto-optic Kerr effect signals, -V$_{in}$/V$_{out}$ (open markers), and the best-fit curves (black lines) from the thermal model for the (a) [Co/Pt]/FePS₃ and (b) [Co/Pt]/MnPS₃ samples.

**Table 3:** Summary of through-plane and in-plane thermal conductivity of $FePS_3$ and $MnPS_3$

| Material | Thermal conductivity at RT ($Wm^{-1}K^{-1}$) | | | Remarks | Ref |
|---|---|---|---|---|---|
| | $k$ | $k_\perp$ | $k_\parallel$ | | |
| $FePS_3$ | 1.35±0.32 | 0.85±0.15 | 2.7±0.3 | Optothermal Raman and TR-MOKE (T=300 K) | [this work] |
| $MnPS_3$ | – | 1.1±0.2 | 6.3±1.7 | TR-MOKE (T=300 K) | [this work] |
| $ZnPSe_3$ | 0.17 | | | Monolayer-theory | [19] |
| $1T$-$TaS_2$ | 4.52 | | | At 300K | [47,48] |
| $2H$-$TaSe_2$ | 14.8 | | | At 270K | [47] |
| $2H$-$TaSe_2$ | 16 | | | Raman | [49] |
| $2H$-$TaSe_2$ | 9 | | | Thin film, Raman | [49] |
| $HfTe_5$ | 6 | | | parallel thermal conductance technique | [50] |
| $MoS_2$ | | | 82 | Time-domain thermoreflectance | [51] |
| $MoS_2$ | | | 16 | Steady-state heater setup at 300K | [52] |
| $MoS_2$ | | | 52 | Few layers, Raman | [53] |
| $MoS_2$ | 55±20 | | 84±17 (300K) | Monolayer, Raman | [54] |
| $MoS_2$ | 35±7 | | 77±25 (300K) | Bilayer, Raman | [54] |
| $MoS_2$ | | | 34.5±4 | Monolayer | [39] |
| $MoS_2$ | 1.76 | | | Pristine, laser flash, at 300K | [55] |
| $MoSe_2$ | | | 35 | Time-domain thermoreflectance | [51] |
| $MoSe_2$ | 42±13 | | | Bilayer | [54] |
| $MoSe_2$ | 24±11 | | 59±18 (300K) | Monolayer, Raman | [54] |
| $MoSe_2$ | 17±4 | | 42±13 (300K) | Bilayer, Raman | [54] |
| $MoSe_2$ | 1.12 | | | Pristine, laser flash, at 300K | [55] |
| $NbSe_3$ | 0.21 | | | | [56] |
| $Ta_2Pd_3Se_8$ | 12.6 | | | Thermal bridge at 350K | [57] |
| $TiS_2$ | 3.45 | | | Laser flash, at 300K | [58] |
| $TiS_3$ | 3.5 | | | Laser flash at 325K | [59] |
| $TiS_3$ | 3.7 | | | Laser flash at 300K | [60] |
| $WS_2$ | | | 120 | Time-domain thermoreflectance | [51] |
| $WS_2$ | | | 32 | Monolayer, Raman at 300K | [61] |
| $WS_2$ | | | 53 | Bilayer, Raman at 300K | [61] |
| $WS_2$ | 2.2 | | | Pristine, laser flash, at 300K | [55] |
| $WSe_2$ | | | 42 | Time-domain thermoreflectance | [51] |
| $WSe_2$ | 0.85 | | | Pristine, laser flash, at 300K | [55] |
| $WSe_2$ | | | 1.2-1.6 | Thin film, at 300K | [62] |
| $ZrTe_3$ | 7-7.5 (±5%) | | | Polycrystalline, four terminal method, at 300K | [63] |
| $ZrTe_5$ | 11.2 | | | At 300K | [64] |
| $ZrTe_5$ | 8 | | | Parallel thermal conductance technique | [50] |
| $ZrTe_5$ | 2 | | | polycrystalline p-type, PPMS at 300K | [65] |

# ■    CONCLUSIONS

In this study, we investigated the phonon and thermal conductivity properties of high quality single crystals of $FePS_3$ and $MnPS_3$ using Raman spectroscopy, optothermal Raman, and TR-MOKE experimental techniques at room temperature. The Raman spectra were excited with three different laser excitation wavelengths of 325 nm, 488 nm, and 633 nm. The data obtained by UV-Raman revealed unique features, which have not been previously detected with regular visible light Raman spectroscopy. Optothermal Raman technique was used to extract the directional average thermal conductivity of bulk $FePS_3$. The data extraction was based on the results of the frequency changes of the main $A_g$ Raman mode as a function of temperature and excitation laser power. The directional average "bulk" thermal conductivity of $FePS_3$ was determined to be $k = 1.35 \pm 0.32 \ \mathrm{Wm^{-1}K^{-1}}$. This value represents a weighted average of the in-plane and through-plane thermal conductivities of the crystal according to $k = \alpha k_\perp + \beta k_\parallel$ in which $\alpha + \beta = 1$ and $\alpha > \beta$. We conducted TR-MOKE measurements in order to determine the anisotropic thermal properties of both $FePS_3$ and $MnPS_3$ bulk crystals. The experiments revealed in-plane thermal conductivity of $2.7 \pm 0.3 \ \mathrm{Wm^{-1}K^{-1}}$ and $6.3 \pm 1.7 \ \mathrm{Wm^{-1}K^{-1}}$ for $FePS_3$ and $MnPS_3$, respectively. The through-plane thermal conductivity measurements reported $0.85 \pm 0.15 \ \mathrm{Wm^{-1}K^{-1}}$ and $1.1 \pm 0.2 \ \mathrm{Wm^{-1}K^{-1}}$ for $FePS_3$ and $MnPS_3$. Comparing the in-plane and cross-plane thermal conductivity values of $FePS_3$ obtained by TR-MOKE measurements with that of attained via optothermal Raman technique with isotropic assumption, the weighted coefficients of $\alpha$ and $\beta$ are calculated as 0.73 and 0.27, respectively. The obtained results are important for understanding the phonon properties and phonon transport in layered antiferromagnetic semiconductors as well as for their applications in spintronic and caloritronic devices.

■   **METHODS**

**Thermal conductivity extraction via Optothermal Method:** We utilized COMSOL software package in order to solve the steady state Fourier's heat equation $\nabla \cdot (\kappa \nabla T) + q''' = 0$ using finite element method. In this equation, $T$ is the temperature distribution of the system and $q'''$ is the laser heat source which can be defined using a Gaussian distribution function as $q''' = \frac{(1-R)P_D \alpha}{(2\pi\sigma^2)} \exp\left(-\frac{x^2+y^2}{2\sigma^2}\right) \exp(-\alpha|z|)$, respectively. Here, $x, y, z$ are the Cartesian coordinates, $P_D$ is the total laser power at the center of the sample's surface ($x = y = z = 0$) and $R$ and $\alpha$ are the sample's reflection and absorption coefficients at the laser excitation wavelength, respectively. The standard deviation of the Gaussian power ($\sigma$) is calculated as $\sigma = r/2$ where $r$ is the radius of the laser spot on sample's surface. The laser spot size should be measured experimentally using knife-edge method[66]. The lateral dimensions and thickness of FePS$_3$ are extracted by careful imaging of the flake using optical and atomic force microscopy (AFM), respectively. The simulated structure from top to bottom corresponds to a 400 nm thick FePS$_3$ flake, 300 nm SiO$_2$ layer, and 4 µm silicon substrate. In order to extract the thermal conductivity, a reiterative procedure was followed. In this model, first a thermal conductivity for the material system is assumed and then the temperature distribution is obtained solving the heat diffusion equation using COMSOL software assuming different values of incident power ($P_D$). At each assumed thermal conductivity value, $\frac{\partial T_{max}}{\partial P_D}$ and then, $\theta = \chi_{\text{Fe}} \frac{\partial T_{max}}{\partial P_D}$ are calculated. The maximum temperature ($T_{max}$) is obtained from the simulations. The temperature coefficient, $\chi_{\text{Fe}}$, was obtained via linear fitting of the experimental Raman peak shift as a function of temperature. Finally, the thermal conductivity is plotted as a function of $\theta$ and compared to the experimental value ($\theta_{exp} = d\omega/dP_D$) which is obtained by linear fitting of Raman peak shift as a function of temperature.

■  **ASSOCIED CONTENT**

**Supporting Information**

The supporting information is available free of charge on the ACS Publication website at DOI:
XXXX

> The supporting information includes a detailed description of sample characterizations,
> optical microscopy and atomic force microscopy, temperature dependent Raman data,UV-
> Raman spectroscopy of different flakes of $FePS_3$ and $MnPS_3$, bare silicon, and diamond
> substrates.

■  **AUTHOR INFORMATION**

**Corresponding Authors**

* Email: fkargar@engr.ucr.edu (F.K.);balandin@ece.ucr.edu (A.B.B.).

ORCID

Ece Aytan: 0000-0002-6605-1463

Fariborz Kargar: 0000-0003-2192-2023

Bishwajit Debnath: 0000-0001-5854-3065

Alexander A. Balandin: 0000-0002-9944-7894

**ACKNOWLEDGEMENTS**

The work at UC Riverside was supported as part of the Spins and Heat in Nanoscale Electronic
Systems (SHINES), an Energy Frontier Research Center funded by the U.S. Department of
Energy, Office of Science, Basic Energy Sciences (BES) under Award # SC0012670. The authors
thank D. Yang at UC Riverside Center for Nanoscale Science and Engineering for his help in sample characterization. This work used the Extreme Science and Engineering Discovery Environment (XSEDE)[67], which is supported by National Science Foundation Grant No. ACI-1548562 and allocation ID TG-DMR130081.

**CONTRIBUTIONS**

A.A.B., E.A. and F.K. conceived the idea of the study. A.A.B. coordinated the project and contributed to the experimental and theoretical data analysis; E.A. carried out the visible light Raman measurements and contributed to the data analysis; F.K. coordinated Raman data analysis and performed numerical simulations; S.G. conducted UV-Raman measurements; M.G. and J. L carried out and analyzed the TR-MOKE measurements under the supervision of R.W.; A.S.M. conducted sample characterization; Z.B. contributed to temperature dependent Raman measurements; Y.L. performed the DFT calculations of the vibrational modes. B.D. contributed to numerical modeling and thermal conductivity extraction; R.K.L. supervised the theory and modeling. F.K. and A.A.B. led the manuscript preparation. All authors contributed to writing and editing of the manuscript.

# Supplementary Information

**Phonon and Thermal Properties of Quasi-Two-Dimensional FePS₃ and MnPS₃**

**Antiferromagnetic Semiconductors**

**Fariborz Kargar[1,×,\*], Ece Aytan[1,×], Subhajit Ghosh[1], Jonathan Lee[2], Michael Gomez[2], ,**
**Yuhang Liu[3], Andres Sanchez Magana[1], Zahra Barani Beiranvand[1], Bishwajit Debnath[3],**
**Richard Wilson[2,4], Roger K. Lake[3,4], and Alexander A. Balandin[1,4,\*]**

[1]Phonon Optimized Engineered Materials (POEM) Center, Department of Electrical and

Computer Engineering, University of California, Riverside, California 92521 USA

[2]Mechanical Engineering Department and Materials Science and Engineering Program,

University of California, Riverside, California 92521 USA

[3]Laboratory for Terascale and Terahertz Electronics (LATTE), Department of Electrical and

Computer Engineering, University of California, Riverside, California 92521 USA

[4]Spins and Heat in Nanoscale Electronic Systems (SHINES) Center, University of California,

Riverside, CA 92521 USA

× Contributed equally to the work.
\* Corresponding authors (A.A.B.): balandin@ece.ucr.edu and (F.K): fkargar@engr.ucr.edu;
web-site: http://balandingroup.ucr.edu/

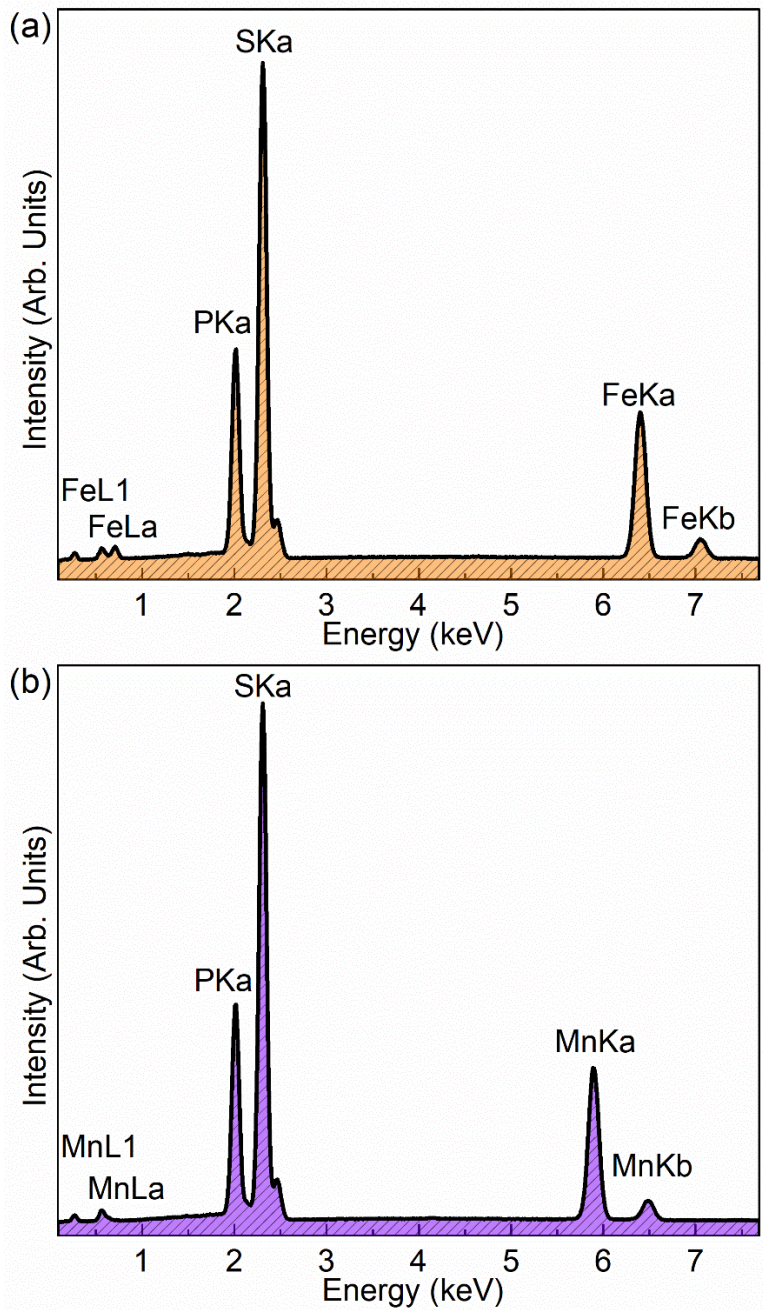

**Figure S1:** The energy-dispersive X-ray (EDX) spectroscopy spectra of (a) FePS3 and (b) MnPS3 confirming the elemental composition of the crystals.

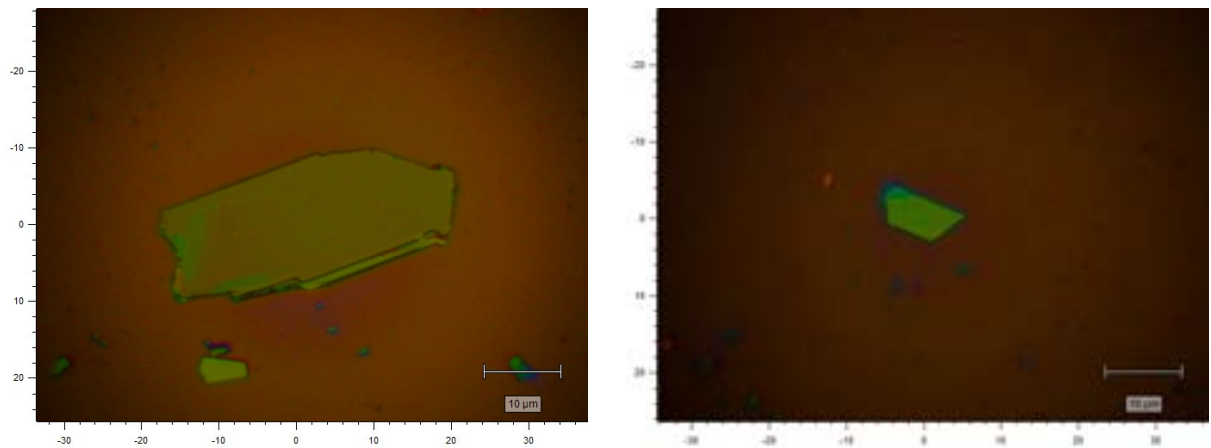

**Figure S2:** Optical images of (a) FePS$_3$ and (b) MnPS$_3$ flakes used for Raman spectroscopy and Raman optothermal measurements.

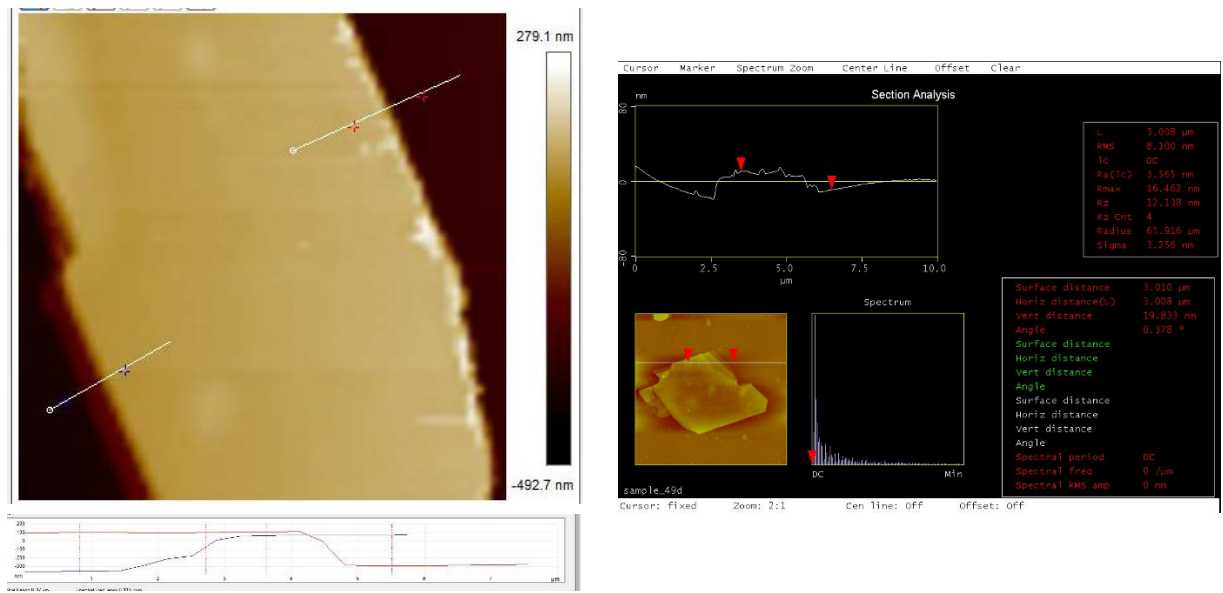

**Figure S3:** Atomic force microscopy (AFM) images of (a) FePS$_3$ and (b) MnPS$_3$ flakes used for Raman spectroscopy and Raman optothermal measurements.

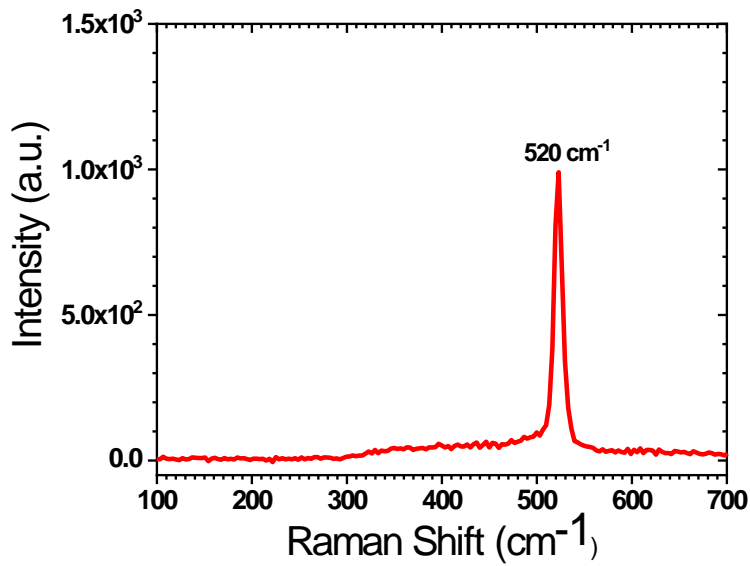

**Figure S4:** UV-Raman spectroscopy of bare silicon substrate

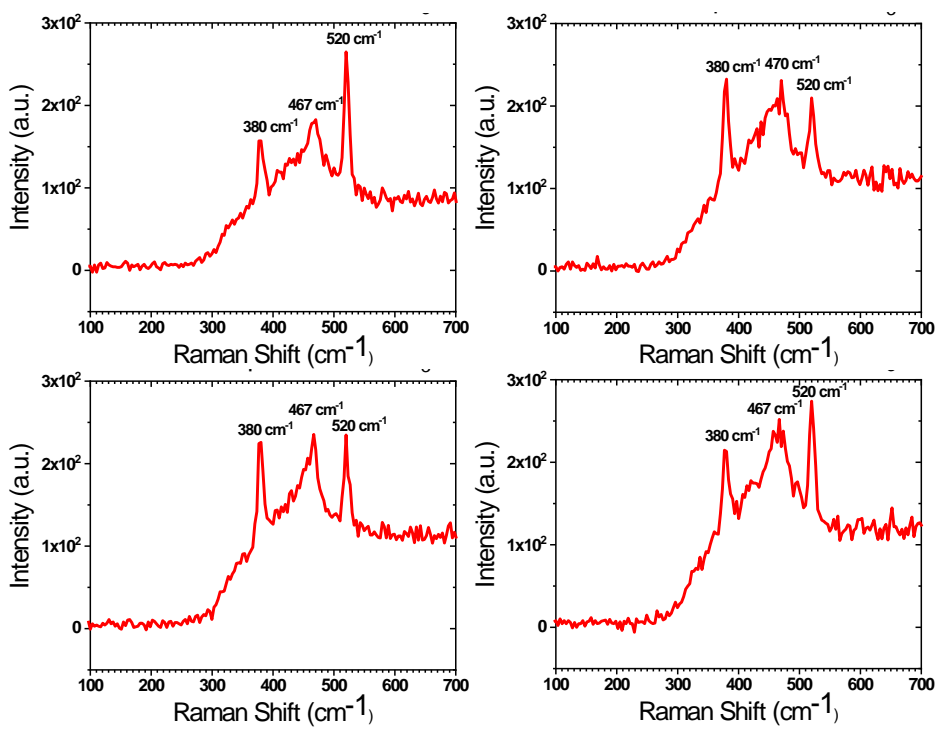

**Figure S5**: UV-Raman spectroscopy on different spots and flakes of FePS$_3$ on silicon substrate.

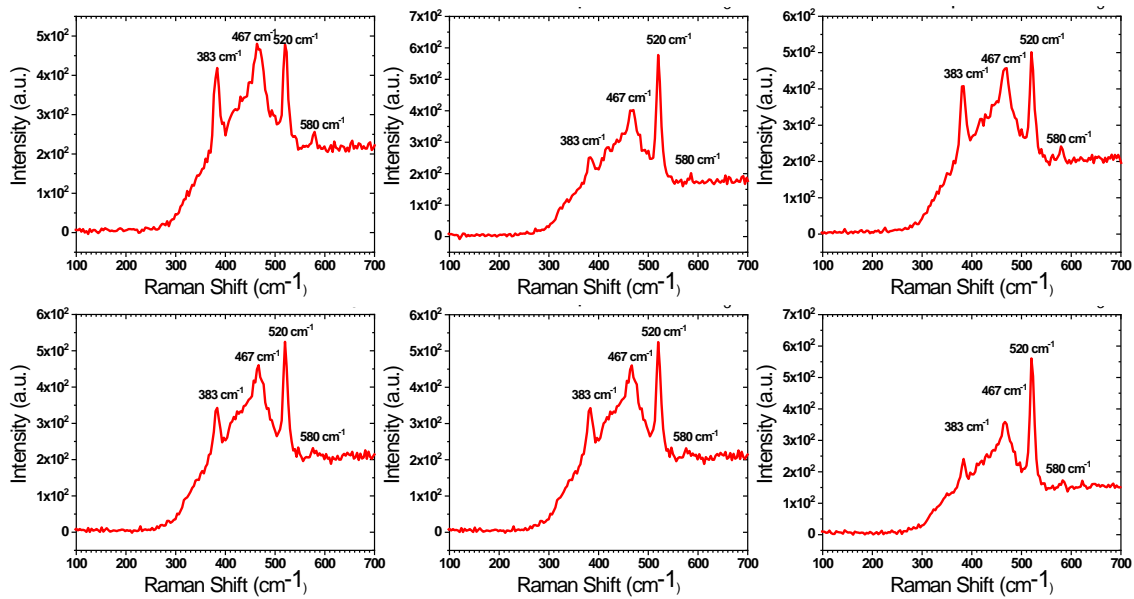

**Figure S6**: UV-Raman spectroscopy on different spots and flakes of MnPS$_3$ on silicon substrate.

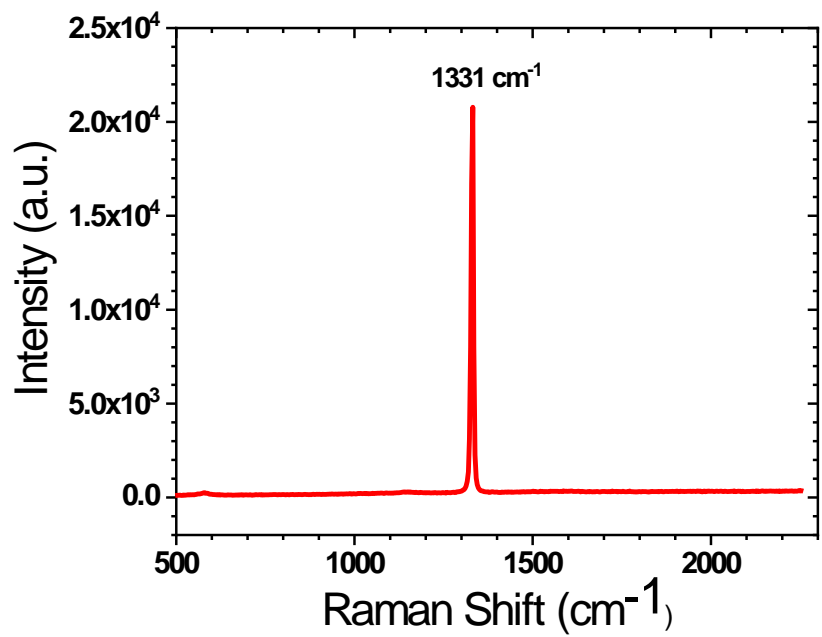

**Figure S7:** UV-Raman spectroscopy of diamond substrate.

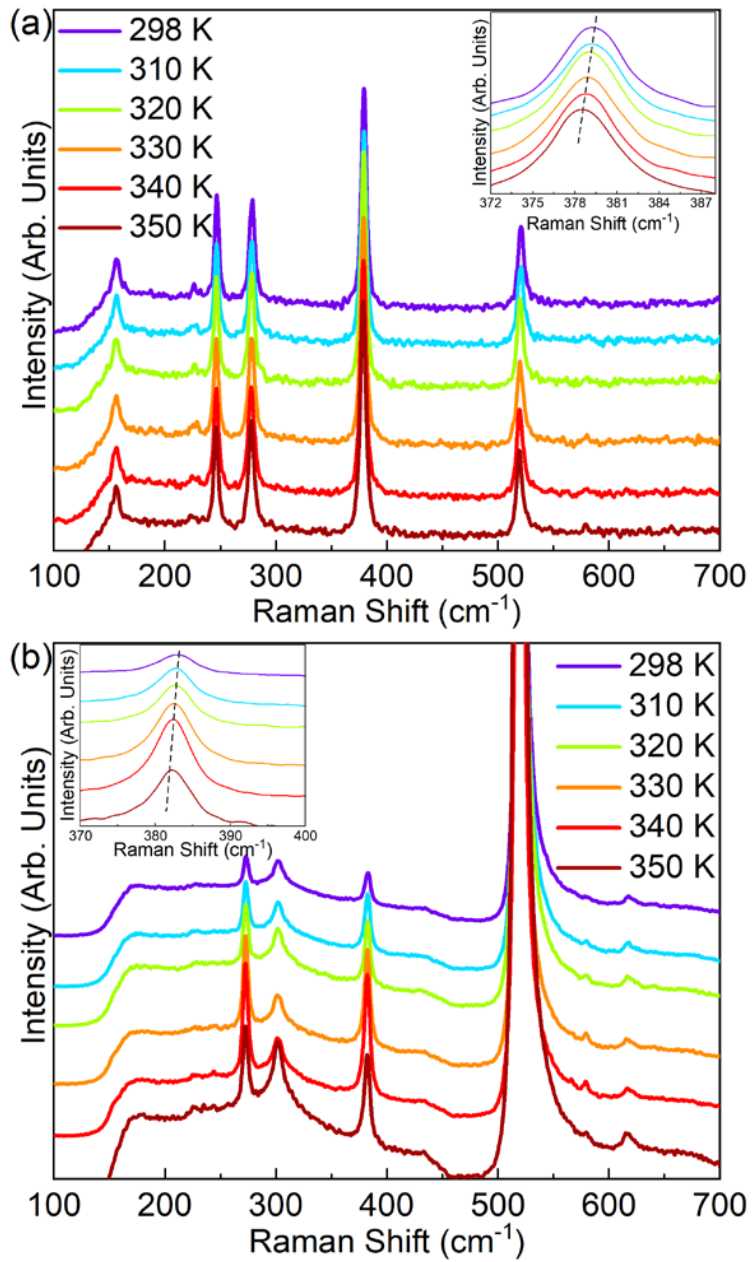

**Figure S8:** Raman spectroscopy of (a) FePS$_3$ and (b) MnPS$_3$ single crystals at room temperature.